\documentclass[12pt]{article}
\usepackage{amssymb}
\usepackage{amsmath}

\begin{document}
\begin{titlepage}
  \begin{flushright}
    hep-th/0512249\\[-1mm]
    TU-762\\[-1mm]
    HD-THEP-06-4
  \end{flushright}
  \begin{center}
    \vspace*{1.4cm}

{\LARGE\bf Gravitational Kaluza-Klein Modes\\[3mm]
 in Warped Superstring Compactification}\vspace{10mm}

{\large
\textbf{Tatsuya Noguchi}\,\footnote{\,noguchi@tphys.uni-heidelberg.de}\\[2mm]
\textit{Institute for Theoretical Physics, Heidelberg University,}\\[1mm]
\textit{D-69120 Heidelberg, Germany} \\[7mm]
\textbf{Masahiro Yamaguchi}\,\footnote{\,yama@tuhep.phys.tohoku.ac.jp}
 and 
\textbf{Masakazu Yamashita}\,\footnote{\,yamasita@tuhep.phys.tohoku.ac.jp}\\[2mm]
\textit{Department of Physics, Tohoku University,}\\[1mm]
\textit{Sendai 980-8578, Japan}
}
\vspace{25mm}

\begin{abstract}
The Kaluza-Klein (KK) modes of graviton are studied in the IIB
superstring compactification where the warped geometry is realized at
the Klebanov-Strassler (KS) throat.  Knowledge of the metric of the KS
throat enables us to determine their wave functions with good
accuracy, without any further specification of the rest of Calabi-Yau
space, owing to the localization of the KK modes.  Mass spectrum and
couplings to the four dimensional fields are computed for some type of
the KK modes, and compared to those of the well-known Randall-Sundrum
model.  We find that the properties of the KK modes of the two models
are very different in both the masses and the couplings, and thus they
are distinguishable from each other experimentally.
\end{abstract}
\vspace{5mm}

\end{center}
\end{titlepage}
\newpage
\setcounter{page}{1}

\section{Introduction}
The warped extra dimension advocated by Randall and Sundrum (RS)
 \cite{RS1,RS2} is a fascinating idea to solve the naturalness problem
 associated with the scale of electro-weak symmetry breaking. In this
 model, the spacetime is a slice of 5-dimensional Anti-de Sitter
 spacetime with two boundaries of four dimensions.  The standard model
 (SM) is assumed to be confined on one of the boundaries, called the
 infra-red (IR) brane.  The characteristic scale on the IR brane is
 re-normalized down to the TeV scale with an appropriate choice of the
 warp factor, providing a solution to the naturalness
 problem. Besides, the Kaluza-Klein (KK) modes of graviton are
 localized near the IR brane, and hence their masses and couplings to
 the four dimensional fields are characterized by the TeV scale, with
 interesting phenomenological consequences.

Despite its attractive features, the RS model as it is has some
drawbacks. One point is that the IR brane on which the SM fields are
confined has a negative tension, which makes its physical
interpretation difficult. Another point is that it has a sharp cut at
the IR brane, which is somewhat ad hoc. Related to this, the warp
factor is given by hand, not fixed by some dynamical mechanism (see
however Ref.~\cite{Goldberger:1999uk}).  All these arguments call for
better understanding of the foundation of this fascinating idea.  In
string theories, the warped metric is natural in the presence of
D-branes. In fact the AdS/CFT~\cite{maldacena} correspondence makes
one anticipate stringy realization of the warped extra dimension
\cite{verlinde, verlinde2}.  An explicit realization has been given in
Ref.~\cite{GKP}. The basic ingredient is type IIB superstring theory
compactified on a Calabi-Yau manifold with the Klebanov-Strassler (KS)
throat, {\it i.e.}  a deformed conifold with non-vanishing RR and NS
fluxes present.  The KS throat has asymptotically AdS metric, and
hence the warped geometry is obtained at the KS throat. The complex
moduli are fixed by the fluxes, so is the warp factor. The latter
determines the characteristic scale of the standard model, provided
that the SM sector resides on D-branes put additionally at the apex of
the KS throat.

The purpose of this paper is to investigate some properties of the
warped string compactification described above. In particular, we will
look into the Kaluza-Klein modes of the graviton in this case. As in
the case of the RS model, the wave functions of the KK modes will be
localized on the infra-red side, around the apex of the KS
throat. This allows us to compute the masses and wave functions of the
KK modes in good approximation, without the information of the
complicated metric of the attached Calabi-Yau manifold.

Though the KS geometry has asymptotically AdS metric, its structure in
the far IR side is quite different. Thus one can expect that the
profiles of the KK modes also differ from those of the RS model. As we
will show later, this is indeed the case. We compute the KK modes
which depend only on the radial direction of the deformed conifold and
compare them with those of the RS model. We find that the mass
eigenvalues are very different from each other. Moreover, we compute
the couplings of the KK modes to the four dimensional fields on the IR
brane, which are inferred from the values of the wave functions there.
It is known that in the RS model all KK modes couple to the SM
particles with the same strength. We find it no longer the case in
this string setup.

In addition to the radial modes which we will study in this paper,
there are modes which depend also on angular directions. The angular
dependence of the KS geometry is somewhat involved, which makes
difficult the study of such modes. We shall make a brief comment on
it, leaving the detailed study of the angular modes for future
investigation.

The organization of the paper is as follows. In the next section, we
study the KK modes of the warped string compactification with the KS
throat. We will solve the wave equations for the KK modes numerically,
and obtain their mass eigenvalues and their couplings to the four
dimensional fields, read from a value of the wave function at the IR
end, for the first several states.  In section 3, after reviewing the
properties of the KK modes of the RS model, we make a comparison of
the two models. A brief comment on the angular dependent modes is made
in section 4. The final section is devoted to conclusions.

\section{KK spectrum and couplings in warped superstring compactification}
\hspace{5mm} 
We consider the warped compactification in superstring theories,
realized on the basis of a deformed conifold \cite{GKP}.  The
background metric of the deformed conifold in the presence of fluxes
for type IIB supergravity was solved by Klebanov and Strassler (the KS
geometry) \cite{KS}.  In the region far from the top of the deformed
conifold, the metric approaches the AdS one, and thus one can use the
KS geometry as a part of the warped compactification.  Since the
deformed conifold is non-compact with its UV end open, one has to glue
it to a Calabi-Yau manifold to make the extra space dimensions
compact, which eventually reproduce a finite four dimensional gravity.

Here we study the gravitational KK modes in this setup.  It is known
that wave functions are localized near the apex of the warped throat
like in the case of the RS model, and thus remain unaffected by the UV
structure, {\it i.e.} the CY manifold glued into the deformed conifold.
Therefore, we perform the calculation of the gravitational KK modes
simply on the KS geometry itself, without any further
compactification.

\subsection{KS solution in the Type IIB supergravity}
\hspace{5mm} 
We begin by briefly reviewing the type IIB supergravity solution on a
deformed conifold which we call the KS solution \cite{KS}.

The deformed conifold is a 3-dimensional complex space defined by the
equation in $\mathbb{C}^4$,
\begin{equation}
(z^1)^2 + (z^2)^2 + (z^3)^2+ (z^4)^2 = \epsilon^2.
\end{equation}
This manifold is a cone, whose base space is topologically equivalent
to $\mathbf{S}^3 \times \mathbf{S}^2$ \cite{candelas}.  In contrast to
the singular conifold ($\epsilon=0$), the base space of the deformed
conifold does not shrink to a point even at $\tau = 0$: it turns out
the (squashed) $\mathbf{S}^3$.  We would like to note that this
non-compact manifold should be regarded as a local description of some
Calabi-Yau manifold, where the large $\tau$ region of the conifold is
glued to a compact CY manifold.

In Ref.~\cite{KS}, a non-singular supergravity solution of Type IIB
superstring theory compactified on the deformed conifold was given,
with $M$ fractional D3-branes at the apex:
\begin{equation}
M = \int_{\mathbf{S}^3} \widetilde{F}_{(3)}.
\end{equation}
The fractional D3-branes, defined as D5-branes wrapped around the
collapsed 2-cycles $\mathbf{S}^2$, serve as sources for the magnetic
RR 3-form flux over $S^3$, and make the 2-form field $B_{(2)}$
dependent on the radial coordinate.

The solution of the type IIB supergravity on the deformed conifold is
given as follows under the assumption that the fractional 3-branes are
smeared over the $S^3$ at $\tau = 0$,
\begin{eqnarray}
 ds_{(10)}^2  &=& h^{-1/2}(\tau) \eta_{\mu \nu} dx^{\mu}dx^{\nu} 
             + h^{1/2}(\tau) ds_{\text{deformed}}^2,\\
 ds_{ \text{deformed} }^2 &=& \frac{1}{2} \epsilon^{4/3} K(\tau) 
    \times \biggl[ \frac{1}{3 K^3 (\tau)} 
           \left[ d \tau^2 + (g^5)^2 \right] 
            + \cosh^2 \left( \frac{\tau}{2} \right) 
           \left[ (g^3)^2 + (g^4)^2 \right] \notag\biggr.\\
&&\biggl. + \sinh^2 \left( \frac{\tau}{2} \right) 
          \left[ (g^1)^2 + (g^2)^2 \right] \biggr], \label{KS} 
\end{eqnarray}
      where 
\begin{eqnarray}
K(\tau) &=& \frac{(\sinh(2 \tau) - 2 \tau)^{1/3}}{2 ^{1/3} \sinh \tau },\\
h(\tau) &=& 2^{2/3} \epsilon^{-8/3} \left( g_{ \text{st} } M {\alpha}' 
\right)^2 I(\tau),\\
I(\tau) &=& \int_{ \tau }^{\infty} dx \frac{x \coth x - 1}{\sinh^2 x}
\left( \sinh 2 x - 2 x\right)^{1/3}.
\end{eqnarray}
       
Note that in the asymptotic region with a large $\tau$ this metric
 becomes the AdS one:
\begin{eqnarray}
ds_{(10)}^2 &\sim&
\frac{R^2}{U^2}ds^2_{(4)} 
+ \frac{U^2}{R^2}\left[ dU^2 + U^2 ds^2_{\mathbb{T}^{1,1}} \right], 
\end{eqnarray}
where
\begin{eqnarray}
U^3 &=& \left( \frac{3}{2} \right)^{3/2} \epsilon^{2} \cosh \tau.
\end{eqnarray}
In terms of coordinate 1-forms $g^i \ (i= 1,\cdots 5)$ dependent on
the coordinates of the base space $\{\theta^{1,2}, \phi^{1,2}, \psi
\}$, the 3-form fluxes and the 5-form fluxes satisfying the equation
of motion are formulated as follows
\begin{eqnarray}
g_{\text{st}} \widetilde{F}_{(3)} &=& \frac{g_{\text{st}}M{\alpha}'}{2} 
        \biggl[ g^5 \wedge g^3 \wedge g^4 (1 - F(\tau))
              + g^5 \wedge g^1 \wedge g^2 F(\tau)\biggr.\notag\\
        && \qquad\qquad\biggl. + F'(\tau) d\tau\wedge 
              (g^1 \wedge g^3 + g^2 \wedge g^4)\biggr],\\
H_{(3)} &=& \frac{g_{ \text{st} } M { \alpha }^{'} }{2} 
             \biggl[d\tau \wedge \left(f'(\tau) g^1 \wedge g^2  
                    + k'(\tau) g^3 \wedge g^4 \right)\biggr.\notag\\
        && \qquad\qquad\biggl. +  
                 \frac{1}{2} 
                 \left(k(\tau) - f(\tau)\right) 
                        g^5 \wedge 
                       (g^1 \wedge g^3 + g^2 \wedge g^4)\biggr],\\
g_{\text{st}} \widetilde{F}_{(5)}
&=& B_{(2)} \wedge \widetilde{F}_{(3)}\notag\\
&=& \frac{\left(g_{ \text{st}} M{\alpha}^{'}\right)^2}{4} 
 l(\tau) g^1 \wedge g^2 \wedge g^3 \wedge g^4 \wedge g^5, 
\end{eqnarray}
where
\begin{eqnarray}
F(\tau) &=& \frac{\sinh \tau - \tau}{2 \sinh \tau }, \\
f(\tau) &=& \frac{\tau \coth \tau - 1}{2 \sinh \tau} (\cosh \tau - 1), \\
k(\tau) &=& \frac{\tau \coth \tau - 1}{2 \sinh \tau} (\cosh \tau + 1),\\
l(\tau) &=& \frac{\tau \coth \tau - 1}{4 \sinh^2
    \tau} (\sinh 2\tau - 2\tau).
\end{eqnarray}

The KS solution should be interpreted as a natural embedding of warped
geometry in string theory. The warp factor at the top of the deformed
conifold is computed in Ref.~\cite{GKP}, with the characteristic mass
scale
\begin{eqnarray}
m_{\text{scalar}} &\sim& M_{\text{Planck}}\exp{\left(-\frac{2 \pi K}{3 M
  g_{\text{st}}}\right)},
\end{eqnarray}
where $M, K$ are RR and NS fluxes respectively. 

\subsection{Numerical results of KK modes} 
We consider the gravitational KK modes in the deformed conifold. Here
we focus on the KK modes whose wave functions depend only on radial
direction of the deformed conifold. This is partly because these
modes would correspond to the KK modes in the RS model, which has a
single extra space dimension.

The Einstein's equation of the type IIB supergravity theory is known
as
\begin{eqnarray}
\mathcal{R}_{MN} &=&  - \frac{1}{8} g_{MN} 
    \left( |H_{(3)}|^2 +  g_{ \text{st} }^{2} | \widetilde{F}_{(3)} |^2 
    \right)\notag\\
&& + \frac{1}{2}\left( 
     \frac{1}{2!} 
     H_{(3)M R_{1}R_{2}}H_{(3)N}^{ \hspace{5.7mm} R_{1}R_{2}}
   + \frac{ g_{ \text{st} }^{2} }{2!} 
     \widetilde{F}_{(3) MR_{1}R_{2}}
     \widetilde{F}_{(3) N}^{ \hspace{5.7mm} R_{1}R_{2}}\right.\notag  \\
&& + \left. 
     \frac{ g_{ \text{st} }^{2} }{2 \cdot 4!} 
     \widetilde{F}_{ (5) M R_{1} R_{2} R_{3} R_{4} }
     \widetilde{F}_{ (5) N }^{ \hspace{5.7mm} R_{1} R_{2} R_{3}
    R_{4}}\right),\label{einstein_01}
\end{eqnarray}
where and hereafter we define that Greek indices run over 4D Minkowski
space, Roman ones over the internal manifold, and capital Roman ones
over 10D spacetime.

The linearized equation with respect to the metric fluctuation
$h_{\mu\nu}(x,\tau)$ turns out to be
\begin{eqnarray}
&&- \frac{1}{2}\nabla^{R}\nabla_{R} h_{\mu \nu}(x, \tau)
 + \frac{1}{2}\nabla^{R}\nabla_{\mu} h_{R \nu}(x, \tau) 
+ \frac{1}{2}\nabla^{R}\nabla_{\nu} h_{R \mu}(x, \tau) \notag \\
&& \hspace{15mm} = - \dfrac{1}{8} h_{\mu \nu}(x, \tau)    
\left(  |H_{(3)}|^2 
        + g_{ \text{st} }^{2} | \widetilde{F}_{(3)} |^2 \right) 
        + \dfrac{g_{ \text{st} }^{2} }{4} 
| \star \mathcal{F}_{(5)} |^2 h_{\mu \nu}(x,\tau),
\end{eqnarray}
where $h_{\mu\nu}(x,\tau)$ is defined as
\begin{eqnarray}
g_{\mu \nu} = h^{-1/2}(\tau) \eta_{\mu \nu} + h_{\mu\nu}(x,\tau).
\end{eqnarray}

Assuming that the fluctuation of metric in the axial gauge would be
  decomposed into the $\tau$-dependent part and a plane wave solution 
  with a definite momentum $k$ as,
\begin{eqnarray}
h_{\alpha\beta}(x^{\mu}, \tau)= \psi(\tau)
 e^{i k \cdot x}\qquad(\alpha,\beta=1,2,3), 
\end{eqnarray}
we find that Eq.~(\ref{KS}) turns out
\begin{eqnarray}
\psi'' + A(\tau) \psi' + B( \tau ) \psi &=& - \frac{ 2^{2/3}
(g_{\text{st}} M \alpha')^2}{ \epsilon^{4/3}} I(\tau) G_{55}(\tau) k^2
\psi,
\end{eqnarray}
where
\begin{eqnarray}
A(\tau) &=& \frac{\partial\ln G_{99}(\tau)}{\partial\tau} +
\frac{\partial\ln G_{77}(\tau)}{\partial\tau} +  \frac{\partial \ln I(\tau)}{\partial\tau},
\end{eqnarray}
and
\begin{eqnarray}
B(\tau) & = & - \frac{1}{4} \cdot \frac{2^{1/3}}{8 I(\tau) G_{77}^2(\tau )}
          \left[ (1-F(\tau))^2  + {k'}^2(\tau) \right] \notag \\
&& - \frac{1}{4} 
          \cdot\frac{2^{1/3}}{16 G_{77}(\tau) G_{99}(\tau)}
          \left[ (k(\tau) - f(\tau))^2 + 4 {F'}^2 (\tau) \right] \notag \\
&& - \frac{1}{4} \cdot \frac{2^{1/3}}{8 I(\tau) G_{99}^2(\tau )}
          ({f'}^2(\tau) +F^2(\tau)) 
          + \frac{1}{4 I(\tau)^2} {I'}^2(\tau) \notag \\
&& - \frac{2^{5/3} l^2(\tau) }{ I^2(\tau) K^4(\tau) 
\left(\cosh^2 \tau -1 \right)^2 }. 
\end{eqnarray}
We adopt here the notation in Ref.~\cite{caceres}, where $G_{ii}\
(i=5,\cdots ,10)$ take the following forms, defined from the KS metric
:
\begin{eqnarray}
&& G_{55}(\tau)\; =\; G_{66}(\tau)\; =\; \frac{1}{6} K^{-2} (\tau), \notag \\
&&G_{77}(\tau)\; =\; G_{88}(\tau)\; =\; \frac{1}{2} K (\tau) \cosh^2 (\tau/2), 
\notag \\
&&G_{99}(\tau)\; =\; G_{1010}(\tau)\; =\; \frac{1}{2} K (\tau) \sinh^2 (\tau/2).
\end{eqnarray}

As for the boundary condition, we require that $d \psi / d \tau$
should be zero at one boundary $\tau = 0$, and that $\psi$ should
approach to $e^{-4 \tau /3 }$ in the large $\tau$ region. The latter
requirement comes from the fact that the background metric approaches
the AdS.

We obtained the mass spectrum for the graviton excitation in the KS
model through the shooting method \cite{ooguri, caceres}.  In
Table~\ref{tab:KS}, we show the mass ratios to the first excited state
and the difference between two neighboring ratios (left), and the
coupling constants between KK modes and the boundary fields (right).
The first KK mode has a mass $4.44 ~ M_{\text{apex}}$, expressed in
the unit
\begin{eqnarray}
M_{\text{apex}} \equiv 
\epsilon^{2/3} / \left( g_{ \text{st} } M {\alpha}' \right).
\end{eqnarray}
The ratios of the coupling constants to the first excited state are
shown, read from the values of the normalized wave functions at the
apex of the warped throat,
\begin{eqnarray}
  g_n \propto \psi(\tau = 0).
\end{eqnarray}

\begin{table}[htbp]
   \centering
   \begin{minipage}{6.0cm}
\renewcommand{\arraystretch}{0.0}
\begin{center}
 \begin{tabular}[h]{|c|c|c|}
\hline
\phantom{.} & & \\
\phantom{.} & & \\
\phantom{.} & & \\
Level & Mass ratio & Difference\\
\phantom{.} & & \\
\phantom{.} & & \\
\phantom{.} & & \\
\hline
\phantom{a} & & \\
6th & 2.48 &    \\
&&0.31\\
5th & 2.17 &    \\
&&0.31\\
4th & 1.86 &    \\
&&0.31\\
3rd & 1.55 &    \\
& & 0.28\\
2nd & 1.27 &    \\ 
& & 0.27\\
1st & 1.00 &    \\
\phantom{a} & & \\
\hline
\end{tabular}
\end{center}
\end{minipage} \hspace{5mm}
   \begin{minipage}{6.0cm}
\renewcommand{\arraystretch}{0.94}
\begin{center}
 \begin{tabular}[h]{|c|c|}
\hline
Level & Coupling ratio\\
\hline
6th & 0.80 \\
5th & 0.89 \\
4th & 0.98 \\
3rd & 1.10 \\
2nd & 1.18 \\ 
1st & 1.00 \\
\hline
\end{tabular}
\end{center}
\end{minipage} 
\caption{
The mass ratios to the first excited state and the
  difference between them (left), and the coupling constants of the KK
  modes to the SM fields localized at the apex of the warped throat,
  normalized by the coupling of the first excited state (right) for the KS 
geometry. They are obtained through the shooting method, where we
evaluated the boundary condition at a point sufficiently near to the
apex, not exactly at the apex, for avoiding computational difficulty.}
\label{tab:KS}
\end{table}

\section{Comparison with Randall-Sundrum Model}
\hspace{5mm}
In this section, we would like to compare the results obtained for the
warped string compactification with those of the Randall-Sundrum
model.

In this model, the spacetime has five dimensions with one extra
dimension compactified on $\mathbf{S}^1 / \mathbb{Z}_2$. The Standard
Model (SM) is thought to be localized on one of two fixed planes under
$\mathbb{Z}_2$ transformation.  Due to the (negative) cosmological
constant $\Lambda( < 0)$, the metric solution of 5D Einstein's
equation is exponentially warped along 5-th dimension $y$:
\begin{equation}
ds^2 = e^{-2 k |y|} \eta_{\mu \nu} dx^{\mu} dx^{\nu} + dy^2,
\end{equation}  
where the curvature constant $k$ is related to $\Lambda$ as
\begin{equation}
k^2 = \frac{- \Lambda}{24 M^3},
\end{equation}
and $M$ is the fundamental mass scale.  It is well known that this
metric could be converted to the AdS one with some coordinate
transformation.  Two 3-branes with opposite tension are required to
locate at each $\mathbb{Z}_2$ fixed point $y=0$ and $r_c$ (the radius
of $\mathbf{S}^{1}$) so that the 5th dimensional space could be
static.  These tensions $T_\pm$ are related to the cosmological
constant through the curvature constant $k$:
\begin{equation}
\frac{T_{+}}{24M^3} = k = - \frac{T_{-}}{24 M^3}.
\end{equation}   

We assume that the SM fields reside on the negative tension 3-brane at
$y=r_{\text{c}}$ (called the IR brane), and obtain the characteristic
scale, which is to be regarded as the mass scale of the Higgs boson
\begin{equation}
m_{\text{Higgs}} = M e^{-k r_c},
\end{equation} 
while the effective 4-dimensional Planck constant
$M_{\text{Pl}}$ is  
\begin{equation}
M_{\text{Pl}}^{(4)} = \frac{M^2}{k}(1- e^{-k r_c} ) \sim M.
\end{equation}

It follows from these relations that even with a moderate choice of
$r_{\text{c}}$ and $k$, one can reproduce an extremely small mass
scale compared with $M$.
 
Next, we discuss the gravitational KK modes.  We define the
gravitational fluctuation $h_{\mu \nu}$ as in Ref.~\cite{RS2}
\begin{equation}
g_{\mu \nu} = e^{- 2 k |y|} \eta_{\mu \nu} + h_{\mu \nu}.
\end{equation}
The linearized Einstein's equation for $h_{\mu \nu}$ is described in
the transverse and traceless gauge
\begin{equation}
\left[ -\frac{1}{2}\partial^2_y 
+ 2 k^2 - 2 k \delta (y) + 2 k \delta (y - y_c) \right] \psi (y)
= \frac{m^2}{2} e^{2 k |y|} \psi (y),
\end{equation}
where $\psi(y)$ is defined as $h_{\mu \nu} = \psi (y) e^{i k \cdot
x}$.  Note that the $\delta$-functions come from the positive and
negative 3-branes.  With the introduction of a new coordinate and a
new wave function $z \equiv \text{sgn}(y)(e^{k|y|}-1)/k$ and
$\widehat{\psi} \equiv \psi e^{k|y|/2}$ , this equation becomes the
Schr\"odinger type with a constant mass term,
\begin{eqnarray}
\left[ -\frac{1}{2} \partial^2_z 
+ \frac{15k^2}{8(k|z|+1)^2} - \frac{3k}{2} \delta(z)
+ \frac{3k}{2(k z_{\text{c}}+1)^2} \delta(z - z_{\text{c}}) \right] 
\widehat{\psi}(z) &=& m^2 \widehat{\psi}(z), \label{RSeigen}
\end{eqnarray}
and the metric becomes conformal
\begin{equation}
ds^2 = \frac{1}{(k|z|+1)^{2}} \left[ \eta_{\mu \nu}dx^{\mu}dx^{\nu} + dz^2 \right].
\end{equation}  

The boundary condition on $\widehat{\psi}(z)$ must be read from the
  condition that all the $\delta$-functions should disappear in
  Eq.~(\ref{RSeigen}) at $z=0_+$ and $z=z_{c-}$, where $a_\pm = a\pm
  \varepsilon$ with a infinitely small positive parameter
  $\varepsilon$,
\begin{eqnarray}
&& \left[ \partial_z \widehat{\psi}(z) \right]_{z = 0+} 
\,\,\,=\; \left[ - \text{sgn}(z) \frac{ 3k }{2(k |z| + 1)} 
   \widehat{\psi}(z) \right]_{z = 0+}, \notag \\
&& \left[ \partial_z \widehat{\psi}(z) \right]_{z = z_{\text{c}}-}
\,=\; \left[ + \text{sgn}(z-z_{\text{c}}) 
  \frac{ 3k }{2(k \zeta(z) + 1)} 
  \widehat{\psi}(z) \right]_{z = z_{\text{c}}-}, 
\end{eqnarray}
where $\zeta(z) \equiv -|z-z_{\text{c}}|+z_{\text{c}}$.

In Table 2, we show the mass spectrum for the KK-modes of the RS model
with $k = 1$ and $y_{\rm{c}} = 6.64$, 
numerically computed through the shooting method \cite{ooguri, caceres}, where the
masses lie at almost the same intervals.\footnote{The IR cutoff $y_c=6.64$ is corresponding to that of the KS calculation in the previous section.}   
The table also shows that the KK modes couple with the field localized
at $z = z_{\rm{c}}$ with almost the same coupling constant, whose difference are of order $0.001\%$. This universality is a characteristic feature of RS
model \cite{davoudiasl}. The authors discussed this
coupling constant universality analytically,
under the assumption of $m_n/k \ll 1$, which is naturally satisfied
when we set the value $ky_c$ for solving the hierarchy problem. 

Our numerical observation of the coupling constant universality here
for RS model would be important when we estimate the numerical error
of the result in the previous section for the KS model, which might
potentially accumulate in the far UV region:
one could estimate it to be somehow similar to that of RS model, i.e., 
0.001\%, because they have the same asymptotic behavior in the UV region.

\begin{table}[htbp]
   \centering
   \begin{minipage}{6.0cm}
\renewcommand{\arraystretch}{0.0}
\begin{center}
 \begin{tabular}[h]{|c|c|c|}
\hline
\phantom{.} & & \\
\phantom{.} & & \\
\phantom{.} & & \\
Level & Mass ratio & Difference\\
\phantom{.} & & \\
\phantom{.} & & \\
\phantom{.} & & \\
\hline
\phantom{a} & & \\
6th & 5.12 &    \\
&&    0.82\\
5th & 4.30 &    \\
&&    0.82\\
4th & 3.48 &    \\
&&    0.82\\
3rd & 2.66 &    \\
& &   0.82\\
2nd & 1.83 &    \\ 
& &   0.83\\
1st & 1.00  &    \\
\phantom{a} & & \\
\hline
\end{tabular}
\end{center}
\end{minipage} 
\hspace{5mm}
\begin{minipage}{6.0cm}
\renewcommand{\arraystretch}{0.0}
\renewcommand{\arraystretch}{0.94}
\begin{center}
 \begin{tabular}[h]{|c|c|}
\hline
Level & Coupling ratio\\
\hline
6th & 1.00 \\
5th & 1.00 \\
4th & 1.00 \\
3rd & 1.00 \\
2nd & 1.00 \\ 
1st & 1.00 \\
\hline
\end{tabular}
\end{center}
\end{minipage} 
\caption{
The mass ratios to the first excited state and the
  difference between them (left), and the coupling
  constants of the KK modes to the SM fields on the IR brane,
  normalized by the first excitation state (right).}
\label{tab:RS}
\end{table}

We are ready to compare the KK spectrum of the warped string
compactification with those of the RS model. Comparison of
Table~\ref{tab:KS} with Table~\ref{tab:RS} shows the clear difference
of the mass spectrum, for example, the ratios of the second KK mode to
the first differ by 67 \%.  In addition to the mass spectra, we would
note that the difference can be seen on the coupling constants: the
warped string compactification based on the KS geometry shows no
longer the coupling constant universality observed in the RS model.

These differences will be a clear target of future collider
experiments such as the Large Hadron Collider (LHC) and a future
lepton collider~\cite{davoudiasl,davoudiasl-2}, where the difference
of the masses and couplings leads to the difference in the loci and
the widths of the s-channel resonances of the KK modes of the
graviton.

At the end of this section, we would like to mention the difference
between the warped string compactification and a slice of AdS${}_5
\times {\cal M}$ \cite{davoudiasl-2}. In the latter case, on top of
the KK modes which are identical to the RS model (a slice of AdS${}_5$),
there are new KK modes whose wave functions depend on coordinates of
${\cal M}$.  On the other hand, the KS geometry, which is used for the
warped string compactification, cannot be decomposed this way.  As we
have seen, the modes with no angular dependence have different mass
eigenvalues and wave functions from the RS model. Furthermore, angular
dependent modes would have more complicated behavior than the case of
the product of the AdS${}_5$ slice and ${\cal M}$, as we will discuss
shortly. Thus, there appears clear difference between the two cases. It is
interesting to investigate whether these two cases can clearly be
distinguished experimentally, which is beyond the scope of the present
paper.

\section{Angular Dependent Modes}
\hspace{5mm}
In this section, we make a brief comment on KK modes
with angular dependence.  As was discussed earlier, the deformed
conifold is a cone whose base manifold is $\mathbf{S}^3 \times
\mathbf{S}^2$. At $\tau=0$ the metric degenerates into that of a round
$\mathbf{S}^3$: 
\begin{eqnarray}
ds_{\text{deformed}}^2 &\sim& \frac{1}{2}
\epsilon^{4/3} \left( \frac{2}{3}\right)^{1/3}
\left[ \frac{1}{2} (g^5)^2 + (g^3)^2 + (g^5)^2 \right],
\end{eqnarray}
and $\mathbf{S}^2$ collapses there. The angular dependent part would
be described by the spherical harmonics $Y_{lm}$ on $\mathbf{S}^2$ and
the harmonics $Z_{pqr}$ on $\mathbf{S}^3$. The modes with $l=0$ do not
vanish at the apex.

A slice of the KS geometry at $\tau \neq 0$ contains the round
$\mathbf{S}^2$ and the squashed $\mathbf{S}^3$. Thus the modes with
the vanishing angular momentum for the round $\mathbf{S}^2$ are
characterized by the isometry of the squashed $\mathbf{S}^3$, {\it
i.e.} $SU(2) \times U(1)$ \cite{shiraishi}. Only these modes can
couple to SM particles when the SM brane is at the top of the deformed
conifold.  On the other hand, if the SM brane is off the top of the
deformed conifold ($\tau \neq 0$), the modes having non-zero angular
momentum associated to the $\mathbf{S}^2$ will have non-vanishing wave
functions at the SM brane, and hence can couple to the SM
particles. Interestingly the strength of the couplings will depend on
the location of the brane off the top. We leave more detailed study on
the angular dependent modes for future study.

\section{Conclusions and Discussion}
\hspace{5mm}
In this paper, we considered the KK modes of the graviton in the
warped string compactification where the warped geometry is realized
at the KS throat. Focusing on the modes with only radial direction
dependence, we computed their masses and coupling constants to the SM
particles which are thought to reside at the top of the deformed
conifold. We compared these results with those of the RS model, and
found clear difference between the two models, which suggests,
interestingly, that that they are distinguishable from each other in
future collider experiments.

Our analysis is based on Ref.~\cite{GKP}, where complex moduli as well
as dilaton are stabilized, while K\"{a}hler moduli remain unfixed. The
issue of the stabilization of the latter moduli was addressed in
Ref. \cite{KKLT}. In the KKLT setup, the K\"ahler moduli are
stabilized by non-perturbative effects such as Euclidean D3-branes and
gaugino condensation on D7-branes, and $\overline{\text{D}}$-branes
are put at the 
apex of the deformed conifold to lift up the otherwise negative vacuum
energy. Here we would like to briefly discuss how they will affect the
results we obtained. As we emphasized in this paper, the
gravitational KK modes have wave functions localized near the apex of
the deformed conifold. The non-perturbative configurations uplifting
the K\"ahler moduli potential, as far as it is added in the UV
Calabi-Yau region, does not affect substantially the KK mode wave
functions. The existence of the $\overline{\text{D}}$-branes opens a new
possibility that the SM fields live on the $\overline{\text{D}}$-branes. However
as far as the low-energy behavior of the KK modes is concerned,
whether the SM lives on the $\overline{\text{D}}$-branes or on the D-branes
makes no difference. We, thus, conclude that the inclusion of the
KKLT setup does not change our results significantly.

Much remains to be seen: firstly, we need to take into consideration
also the modes with angular dependence along the deformed conifold,
and secondly we should investigate experimental signatures of KK modes
in more detail. We leave the study of these issues for future work.

\section*{Acknowledgments}
We would like to thank H. Ishikawa for useful discussions. This work
was supported in part by the Scientific Grants from the Ministry of
Education, Science, Sports, and Culture of Japan, No.~16081202 and
17340062. 

\section*{Note Added}
After completion of this work, we received a preprint
\cite{Firouzjahi:2005qs} which has some overlap with our work.

\end{document}